# Reverse vaccinology in *Plasmodium falciparum 3D7*

Raúl Isea[1*], Rafael Mayo-García[2] y Silvia Restrepo[3]

(1) Fundación IDEA. Hoyo de la Puerta, Baruta, Venezuela.

(2) CIEMAT. Av. Complutense, 40, Madrid 28040, Spain.

(3) Universidad de los Andes, Bogotá, Colombia.

(*) Corresponding author:
Raúl Isea
Fundación IDEA. Hoyo de la Puerta, Baruta, Venezuela.
Email: risea@idea.gob.ve

**SUMMARY**

A timely immunization can be effective against certain diseases and can save thousands of lives. However, for some diseases it has been difficult, so far, to develop an efficient vaccine. Malaria, a tropical disease caused by a parasite of the genus *Plasmodium*, is one example. Bioinformatics has opened the way to new lines of experimental investigation One example is reverse vaccinology that aims to identify antigens that are capable of generating an immune response in a given organism using *in silico* studies. In this study we applied a reverse vaccinology methodology using a bioinformatics pipeline. We obtained 45 potential linear B cells consensus epitopes from the whole genome of *P. falciparum 3D7* that can be used as candidates for malaria vaccines. The direct implication of the results obtained is to open the way to experimentally validate more epitopes to increase the efficiency of the available treatments against malaria and to explore the methodology in other diseases.

**Keywords**
*Reverse vaccinology*, *Plasmodium*, Epitopes, Malaria.

INTRODUCTION

The first vaccine that was successfully tested in history was in 1796 when the 8-year-old child James Phipps was immunized against smallpox by Edward Jenner (1749-1823). He had observed that women who milked cows sometimes had hand injuries and did not develop the disease. After two years and after corroborating the results, Jenner published his findings [1]. He introduced the term *vaccine* derived from the latin word *vaccinus* ("cows"). Two hundred years after the discovery, smallpox is the first disease to be eradicated by humanity.

Although it seemed risky at the time to try an experimental vaccine in a child, there was another famous test of a different vaccine in 1885 by Louis Pasteur (1822-1895). The 9-year-old child Joseph Meister avoided being the victim of rabies after a dog bit him [2]. These two examples show that a timely immunization can be effective against certain diseases.

These two victories are always referenced in the field of vaccinology. However, the idea of immunity was already known from the time of Thucydides (430 B.C.). This was a result of his observations during the epidemic that occurred in Athens probably caused by typhoid fever. Thucydides realized that certain people who contracted the disease did not suffer a relapse although they were in direct contact with infected people [3]. These findings suggested that there might be a possibility of developing an immune response to various diseases.

Bioinformatics [4, 5] has allowed both the development of new computational tools and suggested new lines of experimental investigation. To this must be added the fact that currently it is possible to sequence an entire genome in a matter of hours thanks to the use of solid-state nanopore sequencing as Erika Hayden demonstrated in her work published in Nature in 2012 [6]. According to NCBI data as those of October 2015, almost fourteen thousand genomes have been sequenced, of which 420 are from the *archaea* domain, 7,087 are bacterial, 1,551 are eukaryotic, and 4,845 of the viral type.

Thus, there is a wide range of genetic information that would allow us to identify, for example, those antigens that may be useful for vaccine development, *i.e.* the reverse vaccinology methodology.

Reverse vaccinology

The concept of reverse vaccinology was introduced just a decade ago [7]. Its challenge is to identify those antigens that are capable of generating an immune response in a given organism using *in silico* studies. This new approach stems from the fact that a genome can be visualized as a "catalog" of those antigens that a pathogen can express. Therefore, the genome itself has all the information necessary to generate a vaccine.

The first successful example was against the pathogen *Neisseria meningitidis serogroup B* (hereafter abbreviated MenB) that is commonly known as meningococcus [8, 9], for which a vaccine was obtained from genome analysis after forty years of trials.

The 13 meningococcal serogroups have been identified (abbreviated as A, B, C, D, X, Y, Z, 29E, W135, H, I, K, and L) of which only five can cause epidemics (A, B , C, W135, and Y). Today, vaccines have been implemented for four of them. Unfortunately, the serogroup B one was not effective. This was due to the variation in the sequence of its protein surface and cross-reactivity of the capsular polysaccharide of the serogroup with human tissues [8]. In Venezuela, for example, outbreaks caused by meningococcus in 1998 with 132 cases and 26 deaths, were reported. A year later, the number of cases decreased but with similar number of deaths as the previous year.

The manufacture of a MemB vaccine was possible by sequencing the genome of the strain MC58. Eighteen months after the entry, 570 ORFs (Open Reading Frame) that can be considered potentially antigenic were identified by using bioinformatics tools. Subsequently, some studies in the laboratory were started where only 350 could be expressed in *E.coli*. Once all of them were purified, they were injected into mice in order to examine the antiserum using ELISA and FACS to assess the cellular localization of antigens in meningococcus.

Finally, five proteins (called antigens of Neisseria), having a cross-protection against heterologous strains were obtained. Given the success of being able to obtain a vaccine against MenB, this technique is currently being used on other pathogens such as *Bacillus anthracis* [10], *Streptococcus pneumoniae* [11], *Staphylococcus aureus* [12, 13], *Chlamydia pneumoniae* [14], and *Mycobacterium tuberculosis* [15].

Where do we start?

The first step is to identify the function and location of the genes in the genome. In case they were not listed, both its location and its function should be predicted. To do so, Glimmer [16], Orpheus [17], and ORF Finder [18] applications are mainly used. Then, an alignment of all existing sequences in the genome must be performed. There are various computational methods specifically designed for alignment between pairs of sequences, the best known is Blast (available at http://blast.ncbi.nlm.nih.gov). It uses a heuristic approach and was originally implemented by Altschul *et al* in 1990 [20]. In that sense, the database of the National Center for Biotechnology Information [19] counts on 188,372,017 loci until October 2015 (version 210.0), with a total of 202,237,081,559 of pairs.

At this point, it is also important to consider the scientific computing time required to compare multiple sequences at once. As an example, it should be mentioned that the calculation conducted by the Argonne National Laboratory in which a supercomputer with more than ten thousand processors for simultaneous calculation was used [21]. The aim of the calculation was to identify all the sequences of microbial genomes that were similar to each other and to infer the function of those that were listed as hypothetical or unknown. However, the comparison information generated a volume of several Petabytes of data, i.e. the equivalent of having interconnected twenty five thousand 40 GB hard disks.

In 2011, it was published the excellent performance achieved by the program mpiBLAST supercomputer BlueGene/P for multiple alignments using genes involved in the Influenza A (H1N1) from the NCBI in Proceedings of the European Computing Conference [22]. In this work, those hypothetical genes that so far had not been recorded were identified.

This latest achievement introduces us to the subject of comparative genomics, *i.e.* the area that identifies regions of similarity that may exist between genomes. If all of the genes from different strains of the same species are analyzed, all the resulting information would allow us to identify the antigenic proteins that are similar to each other. With this, a wide range vaccine could be obtained, without restriction of a certain stock or region.

Günter Blobel reported that proteins have intrinsic signals that govern the transportation and their location within the cell [23]. It has also been demonstrated that the role of signal peptides when certain motifs (*i.e.* a very specific combination of amino acids) are required for the secretion of certain proteins. In that sense, Bendtsen *et al.* [24] conducted a review of different computer programs that can predict signal peptides, resulting that the most used program was SignalP [25].

It is also noteworthy to state that the epitopes from an amino acid sequence can be obtained. Several computer programs (for example Antigenic, EMBOSS, ABC-pred, Bcepred, BepiPred, SYFPEITHI and so on) are available. However, the main problem is the large number of false positives that are predicted with them. Such as the case of chromosome 1 of *Plasmodium falciparum* 3D7, where the result indicated of these predictions do not match with the data obtained by experimental methods [26].

Why malaria?

Currently, it is known that malaria is a tropical disease caused by a parasite of the genus *Plasmodium* through the bite of an infected *Anopheles gambiae* mosquito. Out of the 380 species of this type of mosquito, a little over 15% are responsible for transmitting the disease [27]. So far, 172 species of *Plasmodium* are known and of these, only five different strains affect the human species: *P. falciparum, P. vivax, P. ovale, P. malariae* and recently, *P. knowlesi*. In fact, the first recorded case of *P. knowlesi* occurred in 1965 in the United States that was linked to a traveler returning from Malaysia [28]. However, of the five strains mentioned, *P. falciparum* is the most virulent: half of its victims, sadly, are children under five.

Reviewing the scientific literature, it is referenced that in Venezuela in 1854, Luis Daniel Beauperthuy (1807-1881) published the hypothesis that mosquitoes were the probable vectors of malaria and yellow fever in the journal Gaceta de Cumana [29]. Unfortunately, that work was not sufficiently disseminated. In 1880, Charles Louis Alfonso Laveran (1845-1922) discovered a parasite in human blood by examining soldiers in Algeria which called hematozoa of Laveran. Seventeen years later, Ronald Ross (1857-1932) stated that the mosquito is a transmitting agent when he was studying malaria in birds. In1899, the Italian Giovanni Battista Grassi (1854-1925) identified Anopheles as responsible for its transmission [30]. Unfortunately, the work performed in Venezuela by Beauperthuy remains unnoticed.

Is it possible to develop a vaccine against malaria?

To date, there has not been an effective response to this scourge. Ruth Nussenzweig *et al.* published in 1967 a work in which it was showed that it was possible to protect against malaria sporozoites when inoculating *P. berghei* irradiated with X-rays in rodents [31].

It should be also noted the work in which a vaccine in humans was developed and tested in 1987 by the scientific team led by Manuel E. Patarroyo in Colombia. The vaccine, called SPf66, consisted of a mixture of three merozoite antigens [32]. Patarroyo and his collaborators showed an efficacy of 75% of the vaccine at the time of the phase I, while the results in phases II and III ranged between 38% and 60%. Later the Patarroyo team conducted a test in Gambia which unfortunately did not meet their expectations. In the Bolivarian Republic of Venezuela, the team led by Oscar Noya performed similar tests in 1994 and found that SPf66 had a 55% of efficacy against strains of *P. falciparum* [33]. Even today this is the subject of scientific research in Colombia and the rest of the world.

Currently, 17 vaccines are being tested. The biggest hopes are pinned on a vaccine called RTS,S/AS02, which proved to be safe and, in turn, resulted in a decrease in the number of clinical episodes by 30%, according to the trial in Mozambique in 2009 with 2,022 children aged between one and four years. The protection period amounts to 45 months [34].

As indicated above, the feasibility of potential antigens from genomic information to apply new vaccine candidates against malaria will be elucidated, thanks to the success of reverse vaccinology.

MATERIALS AND METHODS

In this paper, the methodology is applied to malaria, but it is possible to used it in other diseases as well. For each protein found in the genome of *P. falciparum 3D7*, three conditions must be met simultaneously for its eventual selection as a potential vaccine candidate. First, those who have at least a transmembrane domain are chosen. Usually the TMHMM program (Transmembrane Helices in Proteins, available at http://www.cbs.dtu.dk/services/TMHMM/) is used [35]. Second, the resulting candidates are evaluated with the SignalP-NN program (available at http://www.cbs.dtu.dk/services/SignalP-2.0/) with the restriction that two threshold values corresponding to MaxS and MeanS equal to 0.82 and 0.52, respectively must be exceeded [25]. The last condition is that the final result should be referred in the database called IEDB (available at http://iedb.org).

Simultaneously, a series of consensus epitopes derived from linear B cells obtained from *in silico* genome-wide analysis of *P. falciparum 3D7* will be selected according to Isea methodology [36-39], it means, all present peptide epitopes are mined in the IEDB database. Subsequently, the Redundancy program available at ExPASy (Protein Analysis Expert System, available at http://expasy.org/tools/redundancy/) is used for discarding those that are repeated. The remaining epitopes are analyzed using the Nomad program (available at http://expasy.org/tools/nomad.html). The latter program performs a local multi-alignment without allowing a gap between different amino acids that make up these epitopes; those compounds are also identified by blocks of 12 amino acids with greater likelihood between them, thanks to the iterative evaluation of an entropy function defined in the work of Hernandez et al. [40]. Lately, the obtained results are evaluated with a program called BepiPred for predicting B-cell epitopes in order to evaluate the final antigenicity [31]. That way, those linear B cells consensus epitopes that can be used to develop vaccines against malaria are obtained.

RESULTS

The prediction of an epitope-based computational vaccine has already provided significant results. In this sense, this methodology is crucial when there is no effective drug available, in which this novel approach regarding epitope prediction for vaccine development is designed. Table 1 shows those proteins which satisfy the three aforementioned conditions set in the materials and methods section on each chromosome, and shows the number of antigenic proteins indicated inside the parenthesis.

Finally, by analyzing all the peptide epitopes present in the IEDB database, hundreds of consensus epitopes are generated after applying the reduction and alignments using respectively the Redundancy and Nomad programs. After evaluating all of them with the BepiPred program, only 45 possible linear B cell epitopes are obtained, which are shown in Table 2. However, it is necessary to implement in vivo studios that allow us to determine the optimal B cell epitopes between 11 to 15.

**Table 1**. Those antigenic proteins that are present in *P. falciparum 3D7* are shown. They should also exhibit at least one transmembrane domain and exceed the threshold values of the SignalP-NN program. Inside the parenthesis the number of antigenic proteins is indicated.

| | |
|---|---|
| Chromosome 1 | PFA0385w (1), PFA0690w (7), PFA0055c (7), PFA0125c (5), PFA0635c (1) |
| Chromosome 2 | PFB0400w (1), PFB0760w (21), PFB0205c (3), PFB0925w (1) |
| Chromosome 3 | PFC0490w (8), PFC0710w (7), PFC0125w (9), PFC0210c (337), PFC0590c (3) PFC0835c (1), PFC0925w (1), PFC0810c (1), PFC1090 (7) |
| Chromosome 4 | PFD0260c (3), PFD0295c (7), PFD0110w (21), PFD0075w (1), PFD0210c (1), PFD0440w (3), PFD0310w (17), PFD0690c (1), PFD0940w (8), PFD0555c (3), PFD1195c (7), PFD1037w (207) |
| Chromosome 5 | PFE0120c (5), PFE0710w (1), PFE1330c (1), PFE1510c (3) |
| Chromosome 6 | PFF0050c (1), PFF0620c (21), PFF0795w (7), PFF0800w (207), PFF0985c (1) PFF1395c (2), PFF0995c (3), PFF1455c (1), PFF1535w (7), PFF1475c (1) |
| Chromosome 7 | MAL7P1.3 (7), PF07_0061 (24), MAL7P1.102 (1), MAL7P1.203 (1), PF07_0100 (1), PF07_0113 (3), MAL7P1.119 (1), MAL7P1.176 (35) |
| Chromosome 8 | PF08_0068 (1), MAL8P1.101 (10), PF08_0063 (214), PF08_0052 (1) PF08_0008 (1), PF08_0050 (3), PF08_0011 (2), PF08_0005 (1) |
| Chromosome 9 | PFI0380c (3), PFI0210c (7), PFI0795w (7), PFI0640c (1), PFI0900w (7) PFI1145w (208), PFI0960w (3), PFI1475w (1817), PFI1730w (11), PFI0920c (1) |
| Chromosome 10 | PF10_0018 (1), PF10_0053 (1), PF10_0166 (1), PF10_0168 (7), PF10_0303 (2), PF10_0313 (3), PF10_0355 (5), PF10_0356 (60), PF10_0392 (1) |
| Chromosome 11 | PF11_0023 (7), PF11_0064 (21), PF11_0224 (13), PF11_0229 (7) PF11_0246 (21), PF11_0256 (7), PF11_0344 (35), PF11_0348 (21) PF11_0381 (214), PF11_0486 (9), PF11_0503 (7) |
| Chromosome 12 | PFL0600w (2), PFL0070c (1), PFL0655w (7), PFL0765w (1), PFL1060c (7), PFL1045w (4), PFL1415w (7), PFL1835w (207) PFL2510w (1), PFL2505c (1) |
| Chromosome 13 | PF13_0066 (3), MAL13P1.60 (13), PF13_0133 (1), MAL13P1.193 (2) MAL13P1.210 (1), MAL13P1.203 (1), PF13_0265 (3), MAL13P1.262 (10) PF13_0277 (1), MAL13P1.285 (1), MAL13P1.410 (207), MAL13P1.319 (3) MAL13P1.342 (1), MAL13P1.465 (7) |
| Chromosome 14 | PF14_0005 (3), PF14_0016 (7), PF14_0045 (1), PF14_0063 (3), PF14_0116 (3), PF14_0199 (21), PF14_0201 (1), PF14_0249 (1), PF14_0250 (1), PF14_0275 (7), PF14_0318 (1), PF14_0342 (1), PF14_0372 (32), PF14_0415 (1), PF14_0440 (1), PF14_0495 (6), PF14_0541 (1) |

**Tabla 2.** Linear B cells consensus epitopes after a full *in silico* analysis of the *P. falciparum 3D7* genome.

| | | |
|---|---|---|
| NGNRHVPNSEDR | QIEEIKKETNQI | TSDLEKKNIPDL |
| AENVKPPKVDPA | QKYSSPSDINAQ | SRNNHPQNTSDS |
| QQQSEKKSISKV | YNNNLERGLGSG | ELDGSKNEWGWS |
| HSNSTTTSLNNN | DLDEPEQFRLPE | ELSSGSSSLEQH |
| EQEIKNKSLEEK | GRNNENRSYNRK | QQPEYEPSVHSI |
| AEQTESPELQSA | LHSDASKNKEKA | LPKREPLDVPDE |
| ELQSAPENKGTG | NRNENNQNTDPY | HNEDVREEIEEQ |
| FPNHLPEELRKQ | LSTNLPYGKTNL | NTSDSQKECTDG |
| SMNNHKDDMNNY | EIKEKKEDLENL | NLEDINKNTRND |
| SPKVLDNERKQS | NAETNPKGKGEV | QHGHMHGSRNNH |
| FDETLGEEDKDL | EEERCLPKREPL | DELDYANDIEKK |
| PLDNIKDNVGKM | SANKPKDELDYA | YIDDVDRDVENY |
| KNQLAGKDEYEA | TLGEEDKDLDEP | ENNDQKKDMEAQ |
| ELRKQTKGVELQ | EWSPCSVTCGKG | EKKNENKNVSNV |
| MPNDPNRNVDEN | KLEETNKEYTNL | SEESHSKTIDPS |

## CONCLUSIONS

By using new computational methodologies that are able to predict new antimalarial vaccines, several advances can be achieved. An outcome could be to extend the period of protection of 45 months obtained with the experimental vaccine RTS,S/AS02.

Epitopes-based vaccines have proven of high utility and can increase the efficiency of the available methods against malaria. To accomplish this work, the selection of those proteins of antigenic importance was restricted according to three basic requirements: a) at least one transmembrane domain must be present; b) values higher than the thresholds set out in the SignalP-NN program must be obtained; and, c) there must be some antigenic evidence published in scientific literature. Thus, about one hundred sequences can be used to develop a vaccine candidate, from over the five thousand ones available throughout the genome of *P. falciparum 3D7* (see Table 1).

After that, 45 potential linear B cells consensus epitopes with an antigenic behavior are obtained with the aforementioned process. Future work would define what criteria should be used to select the length of the consensus epitopes, the analysis of which was restricted to 12 residues, but can range between 11 and 15. Finally experimental tests are necessary to determine the usefulness of this simple computational methodology.   is to open the way to experimentally validate more epitopes as derived vaccines against malaria. Epitopes-based vaccines have proven of high utility and can increase the efficiency of the available methods against malaria.

## CONFLICT OF INTEREST

We have no conflict of interest to declare

## REFERENCES

[1] Jenner, E. (1959) in: Camac CNB, editor. Classics of Medicine and Surgery. New York: Dover; p. 213-40.

[2] Blume, W.E. (1980) *At the Edge of Life: an introduction to viruses*. Washington DC: Public Health Service, National Institutes of Health.

[3] Retief, F.P., and Cilliers, L. (1998). *S. Afr Med. J.* 88(1), 50-53.

[4] Isea, R. (2015). *Global Journal of Advanced Research*. 2(1), 70-73.

[5] Ilzins, O., Isea, R., and Hoebeke, J. (2015). *Open Science Journal of Bioscience and Bioengineering*. 2(5), 60-62.

[6] Hayden, E.C. (2012) *Nanopore genome sequencer makes its debut*. Available at http://www.nature.com/news/nanopore-genome-sequencer-makes-its-debut-1.10051 Nature [10 Octuber 2015]

[7] Rappuoli, R. (2000) *Curr Opin Microbiol*. 3, 445-450.

[8] Granoff, D.M. (2010). *Clin Infect Dis*. 50 (Supplement 2), S54-S65.

[9] DeLisi ,C., Berzofsky, J.A. (1985). *Proc Natl Acad Sci (USA)*. 82(20), 7048-7052.

[10] Ariel, N., Zvi, A., Grosfeld, H., Gat, O., Inbar, Y., *et al*. (2002). *Infect Immun*. 70, 6817-6827.

[11] Wizemann, T.M., Heinrichs, J.H., Adamou, J.E., Erwin, A.L., Kunsch, C., *et al*. (2001). *Infect Immun*. 69, 1593-1598.

[12] Etz, H., Minh, D.B., Henics, T., Dryla, A., Winkler, B., *et al*. 2002. *Proc Natl Acad Sci (USA)*. 99, 6573-6578.

[13] Vytvytska, O., Nagy, E., Blüggel, M., Meyer, H.E., Kurzbauer, R., *et al*. (2002). *Proteomics*. 2, 580-590.

[14] Montigiani, S., Falugi, F., Scarselli, M., Finco, O., Petracca, R., *et al*. (2002). *Infect Immun.* 70, 368-679.

[15] Betts, J.C. (2002). *IUBMB Life*, 53, 239-242.

[16] Delcher, A.L., Harmon, D., Kasif, S., White, O., Salzberg, S.L. (1999). *Nucl Acids Res*. 27, 4636-4641.

[17] Frishman, D., Mironov, A., Mewes, H.-W., Gelfand, M. (1998). *Nucl Acids Res*. 26, 2941-2947.

[18] Rombel, I.T., Sykes, K.F., Rayner, S., Johnston, S.A. (2002). *Gene*. 282, 33-41.

[19] Pruitt, K.D., Tatusova, T., Brown, G.R., Maglott, D.R. (2012). *Nucl. Acids. Res*. 40, D130-D135.

[20] Altschul, S.F., Gish, W., Miller, W., Myers, E.W., Lipman, D.J. (1990). *J. Mol Biol*. 215(3), 403-410.

[21] Balaji, P., Feng, W., Lin, H., Archuleta, J., Matsuoka, S., *et al*. (2010). *Concurrency Computat.: Pract Exper*. 22, 2266-2281.

[22] Borovska, P., Gancheva, V., and Markov, S. (2011). In: Proceedings of the European Computing Conference. Stevens Point (Wisconsin, USA): World Scientific and Engineering Academy and Society (WSEAS); 2011. p. 462-467.

[23] Blobel, G. (2000). *Chembiochem*. 1, 86-102.

[24] Bendtsen, J.D., Jensen, L.J., Blom, N., Heijne, G.V., Brunak, S. (2004). *Protein Eng Des Sel*. 17(4), 349-356.

[25] Bendtsen, J.D., Nielsen, H., Von Heijne, G., Brunak, S. (2004). *J Mol Biol*. 340(4), 783-795.

[26] Isea, R. (2010). *Vaccimonitor*. 19, 15-19.

[27] Paya, A. (2004). *Ginecol Obstet Clínica*. 5(4), 204-210.

[28] Chin, W., Contacos, P.G., Coatney, G.R., Kimball, H.R. (1965). *Science*. 149(3686), 865.

[29] Godoy, G.A., Tarradath, E. (2010). *J Med Biogr*. 18, 38-40.

[30] Capanna, E. (2006). *Microbiol*. 9(1), 69-74.

[31] Ménard, R., Sultán, A.A., Cortés, C., Altszuler, R., Van Dijk, M.R., *et al*. (1997). *Nature*. 385(6614), 336-340.

[32] Patarroyo, M.E., Amador, R., Clavijo, P., Moreno, A., Guzmán, F., *et al*. (1988). *Nature*. 332(6160), 158-161.

[33] Noya, O., Gabaldón, B.Y., Alarcón de Noya, B., Borges, R., Zerpa, N., *et al*. (1994). *J. Infect Dis*. 170(2), 396-402.

[34] Sacarlal, J., Aide, P., Aponte, J.J., Renom, M., Leach, A., *et al*. (2009). *J Infect Dis*. 200, 329-336.

[35] Krogh, A., Larsson, B., Von Heijne, G., Sonnhammer, E.L.L. (2001). *J Mol Biol*. 305(3), 567-380.

[36] Isea, R. (2015). *Vaccimonitor*. 24(2), 93-97.

[37] Isea, R. (2013). *Vaccimonitor*. 22(1), 43-46.

[38] Isea, R. (2013). *Rev. Inst. Nac. Hig. "Rafael Rangel"*. 44 (1), 25-29.

[39] Isea, R. (2013). *Rev. Inst. Nac. Hig. "Rafael Rangel"*. 41 (1), 43-49.

[40] Hernández, D., Gras, R., and Appel, R. (2006). *European Journal of Operational Research*. 185(3), 1276-1284.